\documentclass[aps,prl,reprint,groupedaddress]{revtex4-2}

\usepackage{physics}
\usepackage{amsmath,amsthm,amssymb}
\usepackage{mathrsfs}
\usepackage{graphicx}
\usepackage{subfig}
\usepackage{dcolumn}
\usepackage{bm}

\begin{document}
\title{Rationalizing Unphysical Radiation in the Unruh Effect by Extending Black Hole Spacetime}
	
\author{Yi-Bo Liang}
\author{Hong-Rong Li}
 \email[]{hrli@xjtu.edu.cn}
\affiliation{School of Physics, Xi’an Jiaotong University, Xi’an 710049, China}
	
\date{\today}
\begin{abstract}
	We demonstrate that Schwarzschild spacetime has a conformal extension and that, beyond null infinity, there is a black hole with a timelike singularity.
	In conformal extended spacetime, every null infinity is a killing horizon with vanishing surface gravity.
	When a quantized massless scalar field is taken into this spacetime and different vacuums for the field are defined, thermal radiation coming from the extended black hole could be observed.
	This makes sense, much like the thermal radiation coming from the white hole.
	The Unruh effect is therefore plausible in conformal extended Schwarzschild spacetime.
	It is shown that the thermal radiation coming from past null infinity in Schwarzschild spacetime, which is difficult to imagine as the result of any physical process, is the result of the reduction of the thermal radiation passing through past null infinity in conformal extended spacetime.
	We also present a conformal extension of Kerr spacetime for the first time.
	Then, by examining a quantized massless scalar field on this spacetime, we get the meaningful conclusion that there is thermal radiation coming from a different rotating black hole passing through past null infinity.
	Similarly, the result of the Kerr black hole is consistent with that of the Schwarzschild black hole.
\end{abstract}
	
\maketitle
	A crucial stage in the investigation of the combination of general relativity and quantum theory is the Unruh effect, which is occasionally mistaken for Hawking radiation \cite{hawking1974black} in curved spacetime.
	The Unruh effect is distinct due to part of its thermal radiation, which comes from past null infinity $\mathcal{I}^-$ \cite{kay1991theorems}.
	Despite extensive study on quantum fields in black hole spacetimes \cite{unruh1976notes, candelas1980vacuum, yu2008understanding, unruh1974second, ford1975quantization, ottewill2000renormalized, menezes2018entanglement}, it is still difficult to explain this portion of the Unruh effect in terms of a physical process since $\mathcal{I}^-$ is the ideal boundary of black hole spacetime and it cannot be the source, obviously.
	The strange phenomenon can be understood as, even without accounting for the thermal radiation coming from the past event horizon $\mathcal{H}^-$ that is scattered near $\mathcal{I}^-$, the probability that a two-level system at a low energy level will be excited close to $\mathcal{I}^-$ of the black hole space-time is not zero \cite{yu2008understanding, menezes2018entanglement}.
	This is a challenge in the quantum field theory of curved spacetime up to now.
	For the first time, the source of the thermal radiation coming from $\mathcal{I}^-$ is identified in this article by taking into account the conformal extension of two representative black hole spacetimes and a massless scalar field on them.
	The Unruh effect is therefore in tune in this situation.
	First, contrary to the argument in \cite{penrose1974relativistic}, we find that a conformal extension of Schwarzschild spacetime $\{\bm{M},g_{ab}\}$ exists. 
	The conformal metric $\tilde{g}_{ab}$ on $\bm{M}$ is regular at either $\mathcal{I}^+$ or $\mathcal{I}^-$ in the Eddington coordinate \cite{penrose1974relativistic}.
	Second, extending spacetime beyond $\mathcal{I}^+$ or $\mathcal{I}^-$ can be done as long as the tetrad is provided because the causal structure is determined by the tetrad.
	We show that a timelike singularity exists in the extended area $C$.
	Then, it is simple to quantize a massless scalar field on $\{\bm{M},\tilde{g}_{ab}\}$ because of the well-known methods for doing so on $\{\bm{M},g_{ab}\}$ \cite{fulling1977alternative}.
	The definition of the Hartle-Hawking vacuum \cite{candelas1980vacuum} for the field and the transit of thermal radiation via $\mathcal{I}^-$ are made possible as a result.
	Therefore, the only source of the thermal radiation is $C$, which is located beyond  $\mathcal{I}^-$. 
	Similar to the thermal radiation emanating from the white hole, this makes sense.
	The Unruh effect is undoubtedly reasonable in conformal extended Schwarzschild spacetime.
	If the field is confined on $\{\bm{M},g_{ab}\}$, the Hartle-Hawking vacuum for the confined field can be obtained.
	The meaningful conclusion that follows is that the thermal radiation coming from past null infinity in Schwarzschild spacetime is the result of the reduction of the thermal radiation passing through past null infinity in conformal extended spacetime.
	
	After that, we take into account the Kerr spacetime conformal extension and find more unique rotating black holes $\bm{K^\prime}s$ beyond null infinities.
	A massless scalar field can be introduced and quantized on a globally hyperbolic subspacetime of the conformal extended spacetime.
	CCH and FT vacuums \cite{candelas1981quantization, frolov1989renormalized} can be defined for the field in the absence of stationary Hadamard states \cite{kay1991theorems}.
	We then find that heat radiation is traveling through $\mathcal{I}^-$, and $\bm{K^\prime}$ is its sole source.
	Similarly, the Unruh effect is undoubtedly reasonable in conformal extended Kerr spacetime.
	If the field is confined on a globally hyperbolic subspacetime of Kerr spacetime, the CCH and FT vacuums for the confined field can be obtained.
	Clearly, we arrive at the same conclusion as in the Schwarzchild case.
	
	\begin{figure*}
		\includegraphics[width=12.9cm]{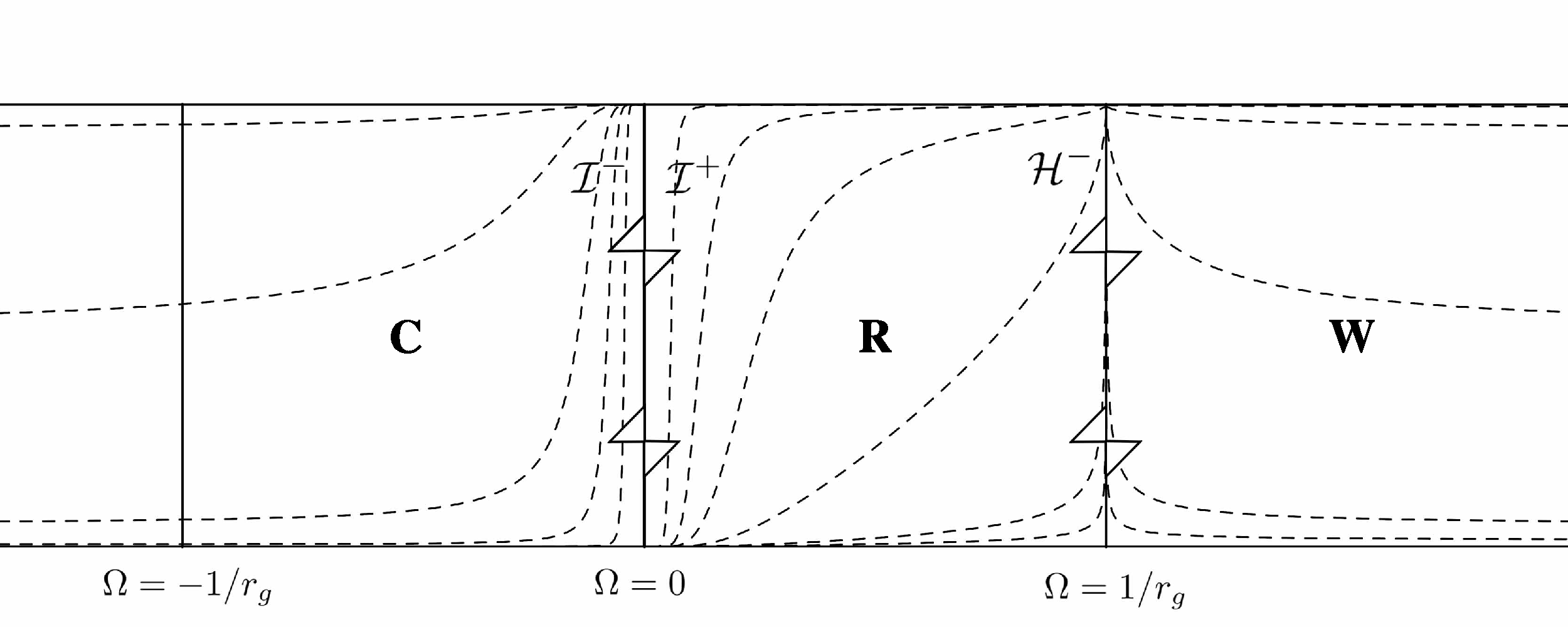}
		\caption{\label{fig:1} The subspacetime is $R\cup W\cup C \cup \mathcal{H}^- \cup\mathcal{I}^+$. Lightcones are drawn on $\mathcal{H}^-$ and $\mathcal{I}^+$. The dashed and solid vertical lines are hypersurfaces of $t=constant$ and $\Omega=constant$ respectively. $\mathcal{I}^+$ of $R$ is $\mathcal{I}^-$ of $C$.}
	\end{figure*}
	Schwarzschild spacetime is $\{\bm{M},g_{ab}\}$ whose metric $g_{ab}$ takes the form,
	\begin{align}
		\mathrm{d}s^2=&-\left(1-\frac{2M}{r}\right)\mathrm{d}t^2+
		\left(1-\frac{2M}{r}\right)^{-1}\mathrm{d}r^2\notag\\
		&+r^2\left(\mathrm{d}\theta^2+\sin^2\theta\mathrm{d}\varphi^2\right),
	\end{align}
	except on the event horizon.
	
	We choose a conformal factor $\Omega=1/r$ and express the conformal metric $\tilde{g}_{ab}=\Omega^2 g_{ab}$ in outer Eddington coordinate \cite{penrose1974relativistic},
	\begin{align}
		\mathrm{d}\tilde{s}^2=&-\left(\Omega^2-2M\Omega^3\right)\mathrm{d}u^2
		+2\mathrm{d}\Omega\mathrm{d}u\notag\\
		&+\mathrm{d}\theta^2+\sin^2\theta\mathrm{d}\varphi^2.
	\end{align}
	
	The region $R$ of $0<\Omega<1/r_g$ can be extended beyond $\Omega=0$ since the metric is regular at $\Omega=0$.
	It is obvious that the area $C$ outside $\Omega=0$ does not resemble $R$ as described in \cite{penrose1974relativistic}.
	There is no longer an event horizon in $C$ since we can see that $r=-1/r_g=-1/2M$ is not the coordinate singularity in $C$.
	As a result, there is a timelike singularity at $r=0$ since the vector $(\partial/\partial u)^a$ is always timelike in $C$.
	The traditional method \cite{frolov2012black} can be used to determine $C's$ conformal structure.
	
	The null hypersurface of $\Omega=0$ is $\mathcal{I}^+$ of $R$.
	$(\partial/\partial u)^a$ is a killing vector field, $k^a$ is normal vector of $\mathcal{I}^+$ and is null there,
	\begin{gather}
		k^a=\left.g^{ab}\nabla_b \Omega\right|_{\Omega=0}=\left.\left(\pderivative{u}\right)^a\right|_{\Omega=0},\\
		g_{ab}k^a k^b=0,\\
		\nabla^a \left(k^b k_b\right)=0.
	\end{gather}
	A killing horizon with vanishing surface gravity is therefore $\mathcal{I}^+$.
	
	On $\mathcal{I}^+$, $-(\partial/\partial \Omega)^a$ is another null vector field.
	With the time orientation specified by the orthonormal tetrad $\{(\bm{e}_I)^a\}$ chosen for simplicity, 
	\begin{gather}
		(\bm{e}_0)^a=\left(\pderivative{u}\right)^a+\frac{1}{2}(b-1)\left(\pderivative{\Omega}\right)^a,\\
		(\bm{e}_1)^a=\left(\pderivative{u}\right)^a+\frac{1}{2}(b+1)\left(\pderivative{\Omega}\right)^a,\\
		(\bm{e}_2)^a=\left(\pderivative{\theta}\right)^a,\quad (\bm{e}_3)^a=\frac{1}{\sin\theta}\left(\pderivative{\varphi}\right)^a,
	\end{gather}
	where
	\begin{gather}
		b=\Omega^2-2M\Omega^3,
	\end{gather}
	it can be seen that the causal structure is determined, and the hypersurface of $\Omega=0$ is $\mathcal{I}^-$ of $C$.
	We can then draw the conclusion that $\mathcal{I}^+$ of $R$ is just $\mathcal{I}^-$ of $C$, see as Fig.~\ref{fig:1} when $\theta$ and $\varphi$ are both constants.
	
	In inner Eddington coordinate with conformal factor $\Omega$, the metric takes the form \cite{penrose1974relativistic},
	\begin{align}
		\mathrm{d}\tilde{s}^2=&-(\Omega^2-2M\Omega^3)\mathrm{d}v^2-2\mathrm{d}\Omega\mathrm{d}v\notag\\
		&+\mathrm{d}\theta^2+\sin^2\theta\mathrm{d}\varphi^2.
	\end{align}
	\begin{figure}[t]
		\includegraphics[width=8.6cm]{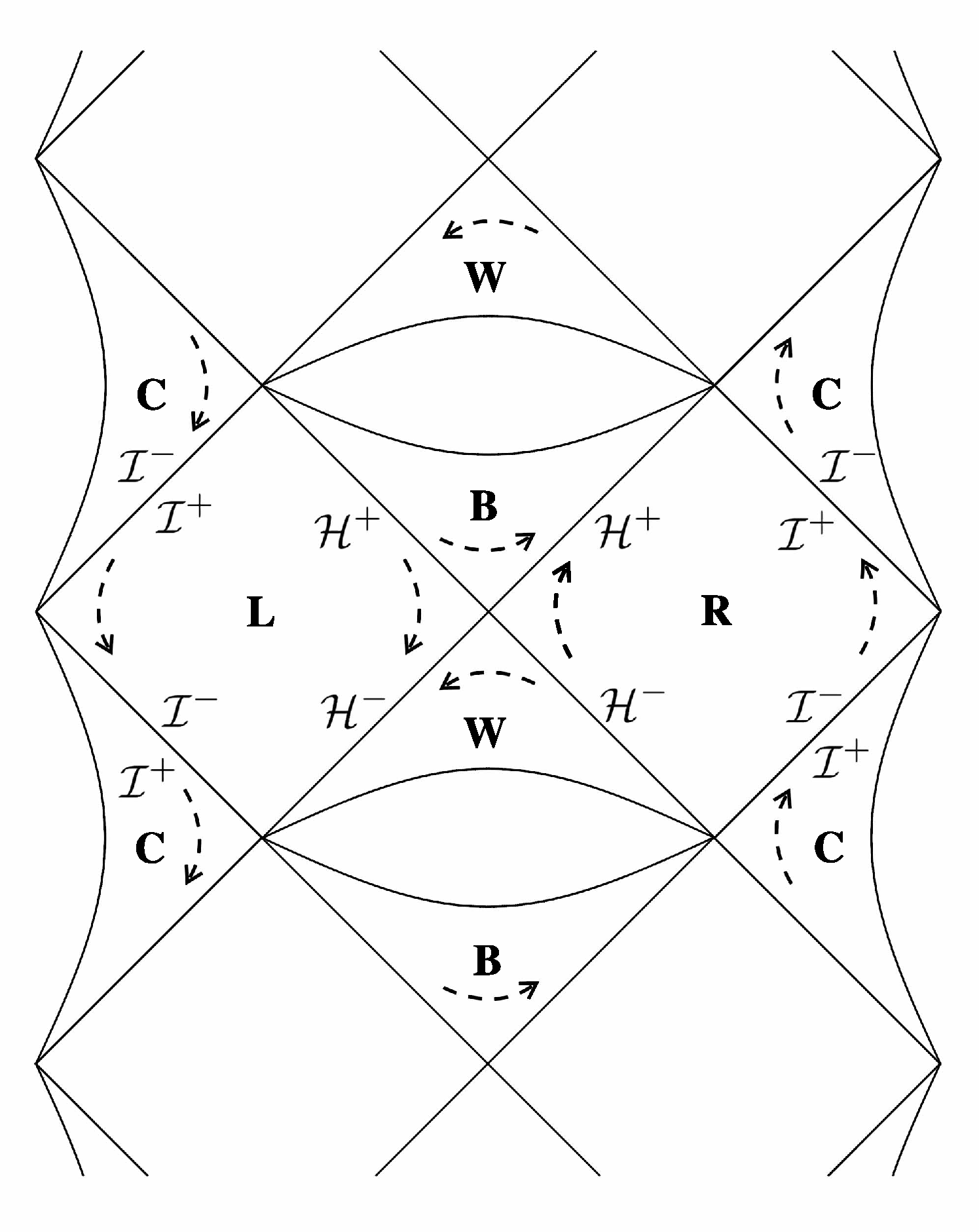}
		\caption{\label{fig:2} Conformal extension of Schwarzchild spacetime goes up and down indefinitely. The dashed arrows represent the killing vector field, which is timelike in $R$ and can be specified consecutively on the whole spacetime. $\bm{M}=R\cup L\cup B\cup W\cup \mathcal{H}^+\cup \mathcal{H}^-$.}
	\end{figure}

	Similarly, we deduce that $\mathcal{I}^-$ of $R$ is just $\mathcal{I}^+$ of $C$ and $\mathcal{I}^-$ is also a killing horizon with vanishing surface gravity.
	There are two separate classes of killing horizons in the conformal extended spacetime, which makes it different.
	
	In Fig.~\ref{fig:2}, we can see $\bm{M}\cup C$ together, but that does not mean we could use a unitary coordinate to describe them together. 
	It resembles the conformal extension obtained in \cite{halavcek2013analytic} partly, but the latter uses a different conformal factor given in \cite{ashtekar1978unified}.
	Killing vector fields on the conformal extended Schwazschild spacetime are distinct from those on the regular Schwarzschild spacetime obtained by solving the singularity \cite{ashtekar2023regular}.
	In various circumstances mentioned above, a larger spacetime is taken into account.
	
	The preceding results show that Fig.~\ref{fig:2}'s representation of the conformal extended Schwarzschild spacetime is just its simplest structure.
	Fortunately, there is a fixed globally hyperbolic subspacetime $\{\bm{M},\tilde{g}_{ab}\}$ with spacelike Cauchy surfaces.
	Then we consider a massless scalar field on $\{\bm{M},\tilde{g}_{ab}\}$.
	Using the definition of $\tilde{\phi}=\Omega^{-1}\phi$, $\tilde{\phi}$ satisfies the wave equation \cite{ashtekar1978unified},
	\begin{equation}
		\tilde{\grad}^a \tilde{\grad}_a \tilde{\phi}-\frac{1}{6}\tilde{R} \tilde{\phi}=0,
	\end{equation}
	$\tilde{\grad}_a$ is the derivative operator compatible with $\tilde{g}_{ab}$.
	
	In the coordinate $\{\zeta,\eta,\theta,\varphi \}$ given in \cite{frolov2012black}, the behavior of the ingoing mode $\overset{\leftarrow}{U}_{\omega lm}= \Omega^{-1}\overset{\leftarrow}{u}_{\omega lm}$ on $\mathcal{H}^+$ of $R$ is
	\begin{equation}
		\overset{\leftarrow}{U}_{\omega lm}\sim e^{(i\omega/\kappa)\ln[\sinh(\tan\eta)]},
	\end{equation}
	(The $lm$ terms have been omitted.)where $\kappa$ is surface gravity of the event horizon and $\overset{\leftarrow}{u}_{\omega lm}$ is given in \cite{candelas1980vacuum}.
	Under wedge reflection isometry $\{\zeta,\eta, \theta, \varphi\}\rightarrow \{-\zeta, -\eta, \theta, \varphi\}$, regions $R$ and $L$ are interchanged.
	$\overset{\leftarrow}{V}^\star_{\omega lm}$ is the reflection of $\overset{\leftarrow}{U}_{\omega lm}$, on $\mathcal{H}^-$ of $L$,
	\begin{equation}
		\overset{\leftarrow}{V}^\star_{\omega lm}\sim 	e^{(i\omega/\kappa)\ln[-\sinh(\tan\eta)]},
	\end{equation}
	
	The linear combinations below will be analytic and purely positive frequency with respect to $\eta$ \cite{wald1994quantum},
	\begin{gather}
		\overset{\leftarrow}{W}_{\omega lm}=A(\overset{\leftarrow}{U}_{\omega lm}+e^{-\pi\omega/\kappa}\overset{\leftarrow}{V}^\star_{\omega lm}),\\
		\overset{\leftarrow}{W}^\prime_{\omega lm}=A(e^{-\pi\omega/\kappa}\overset{\leftarrow}{U}^\star_{\omega lm}+\overset{\leftarrow}{V}_{\omega lm}),\\
		A=\frac{e^{\pi\omega/2\kappa}}{\sqrt{2\sinh(\pi\omega/\kappa)}}.
	\end{gather}
	Similarly, outgoing modes $\overset{\rightarrow}{W}_{\omega,l,m}$ and $\overset{\rightarrow}{W}_{\omega,l,m}$ can be obtained.
	$\overset{\leftarrow}{U}_{\omega lm}$ and $\overset{\rightarrow}{U}_{\omega lm}$ are a complete orthonormal family of complex valued solutions of the wave equation since the Klein-Gordon scalar product is invariant after this conformal transformation.
	
	The quantization of the field can be done as,
	\begin{align}
		\tilde{\phi}=
		&\sum\limits_{l,m}
		\int_{0}^{+\infty}\mathrm{d}\omega
		\bigg[\overset{\rightarrow}{a}_{\omega lm} \overset{\rightarrow}{W}_{\omega l m}
		+\overset{\rightarrow}{a}^\prime_{\omega lm}\overset{\rightarrow}{W}^\prime_{\omega l m}\notag\\
		&+\overset{\leftarrow}{a}_{\omega lm} \overset{\leftarrow}{W}_{\omega l m}
		+\overset{\leftarrow}{a}^\prime_{\omega lm}\overset{\leftarrow}{W}^\prime_{\omega l m}(x)\bigg]+h.c.
	\end{align}
	The operators $\overset{\rightarrow}{a}_{\omega lm}, \overset{\rightarrow}{a}^\prime_{\omega lm}, \overset{\leftarrow}{a}_{\omega lm}$ and $\overset{\leftarrow}{a}^\prime_{\omega lm}$ are interpreted as annihilation operators.
	Then Hartle-Hawking vacuum $\ket{0}$ can be defined on $\{\bm{M},\tilde{g}_{ab}\}$ that for all $\omega,l,m$,
	\begin{equation}
		\overset{\rightarrow}{a}_{\omega l m}\ket{0}=\overset{\rightarrow}{a}^\prime_{\omega l m}\ket{0}=\overset{\leftarrow}{a}_{\omega lm} \ket{0}=\overset{\leftarrow}{a}^\prime_{\omega lm} \ket{0}=0.  
	\end{equation}
	The Wightman function can therefore be obtained.
	By employing the technique described in \cite{benatti2004entanglement, yu2008understanding}, it can be observed that the probability that a static two-level system at ground state will be excited near $\mathcal{I}^-$ is not zero.
	Clearly, $\overset{\leftarrow}{u}_{\omega lm}$ go to zero as they approach $\mathcal{I}^-$, so $\overset{\leftarrow}{U}_{\omega lm}$ do not,
	\begin{equation}
		\overset{\leftarrow}{U}_{\omega l m}\sim
		\begin{cases}
			\ 0,&at \ \mathcal{H}^- of R;\\
			\ \overset{\leftarrow}{B}_{\omega l}e^{-i\omega v},&at \ \mathcal{H}^+ of R;\\
			\ e^{-i\omega v},&at \ \mathcal{I}^- of R;\\
			\ \overset{\leftarrow}{A}_{\omega l}e^{-i\omega u},&at \ \mathcal{I}^+ of R.
		\end{cases}
	\end{equation}
	\begin{figure*}
	\includegraphics[width=12.9cm]{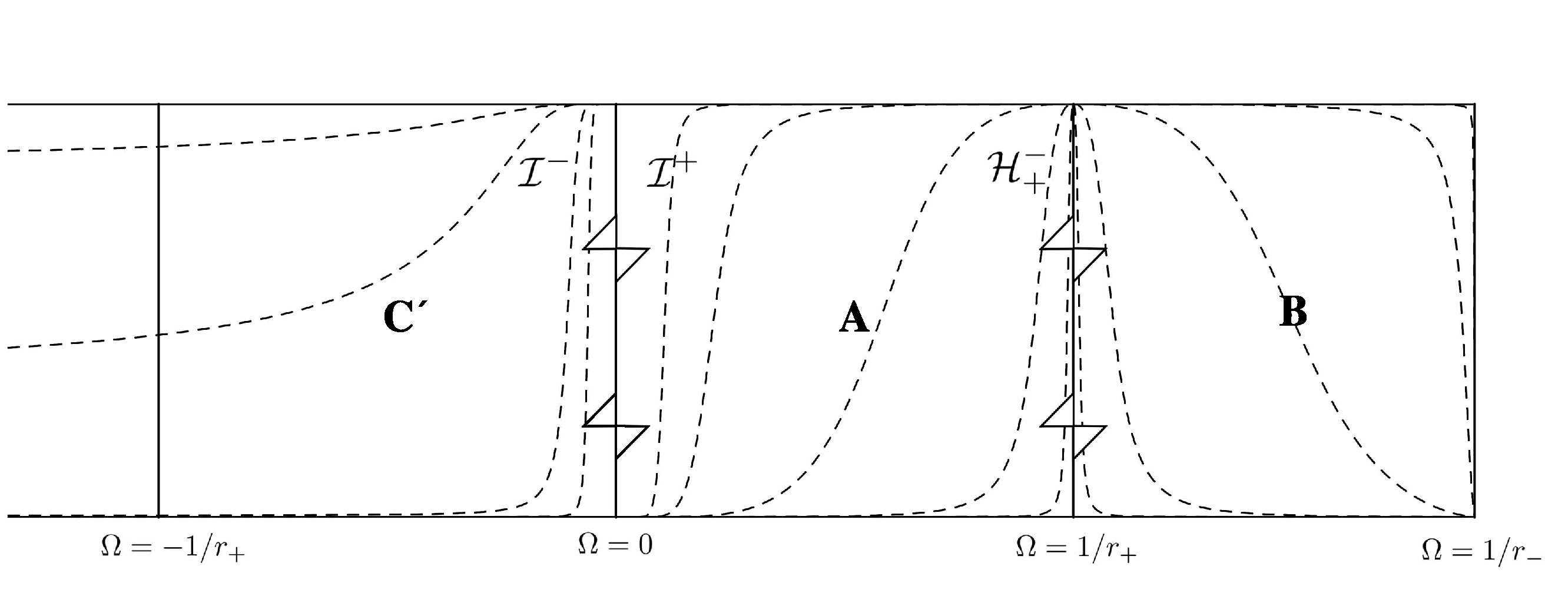}
	\caption{\label{fig:3}
		The subspacetime is $A\cup B\cup C^\prime\cup \mathcal{H}^-_+\cup\mathcal{I}^+$.
		Both $\varphi_+$ and $\theta$ are set constants, and lightcones are drawn on $\mathcal{H}_+^-$ and $\mathcal{I}^+$.
		The dashed and solid vertical lines are hypersurfaces with $t=constant$ and $\Omega=constant$ respectively. $\mathcal{I}^+$ of $A$ is $\mathcal{I}^-$ of $C^\prime$.}
	\end{figure*}

	The important fact indicates that $C$ is emitting thermal radiation that is traveling through $\mathcal{I}^-$.
	This makes sense because it is similar to the thermal radiation coming from the white hole.
	In conformal extended Schwarzschild spacetime, the Unruh effect makes perfect sense, and it is fair to consider this spacetime to be more physical.
	Additionally, by multiplying $\tilde{\phi}$ by $1/r$, we can restrict the quantized field to $\{\bm{M},g_{ab}\}$.
	The vacuum for the confined field is just the usual Hartle-Hawking vacuum. 
	This fact strongly implies a connection between the thermal radiation passing through $\mathcal{I}^-$ in the conformal extended spacetime and the thermal radiation coming from $\mathcal{I}^-$ in Schwarzschild spacetime.

	In Kerr spacetime $\{\bm{K},g_{ab}\}$, we first use the outer Eddington coordinate and choose a conformal factor $\Omega=1/r$.
	The conformal metric, which is regular at the past event horizon $\mathcal{H}_+^-$ and $\mathcal{I}^+$ takes the form \cite{chandrasekhar1985mathematical},
	\begin{align}
		\mathrm{d}\tilde{s}^2=&-\left(\Omega^2-\frac{2M\Omega}{\rho^2}\right)\mathrm{d}u^2\notag\\
		&-\frac{4aM\Omega\sin^2\theta}{\rho^2}\mathrm{d}u\mathrm{d}\varphi_+\notag\\
		&+\left(1+a^2 \Omega^2+\frac{2Ma^2 \Omega\sin^2\theta}{\rho^2}\right)\sin^2\theta\mathrm{d}\varphi_+^2\notag\\
		&-2a\sin^2\theta\mathrm{d}\varphi_+\mathrm{d}\Omega
		+2\mathrm{d}u\mathrm{d}\Omega\notag\\
		&+\Omega^2\rho^2\mathrm{d}\theta^2.
	\end{align}
	
	The region $A$ of $0<\Omega<1/r_+$ can be extended beyond $\Omega=0$ since the metric is regular at $\Omega=0$.
	It is obvious that the area $C^\prime$ outside $\Omega=0$ resembles $C$ which is the region of $-\infty<r<r_-$ in Kerr spacetime.
	However, $C^\prime$ and $C$ are completely distinct and appear complimentary when $\theta=\pi/2$.
	It can be seen that $r=-1/r_-=-1/[M-(M^2-a^2)^{1/2}]$ and $r=-1/r_+=-1/[M+(M^2-a^2)^{1/2}]$ is not the coordinate singularity in $C^\prime$.
	As a result, there is a timelike singularity in $C^\prime$ at $r=0$ when $\theta=\pi/2$.
	
	$(\partial/\partial u)^a$ is a killing vector field, $k^a$ is normal vector of $\mathcal{I}^+$ and is null here,
	\begin{gather}
		k^a=g^{ab}\nabla_b \Omega|_{\Omega=0}=\left.\left(\pderivative{u}\right)^a\right|_{\Omega=0},\\
		g_{ab}k^a k^b=0,\\
		\nabla^a \left(k^b k_b\right)=0,
	\end{gather}
	$\mathcal{I}^+$ is thus a killing horizon with vanishing surface gravity.
	
	On $\mathcal{I}^+$, $(\partial/\partial \Omega)^a$ is another null vector field.
	With the time orientation specified by the ``static'' orthonormal tetrad $\{(\bm{e}_I)^a\}$ chosen for simplicity,
	\begin{gather}
		(\bm{e}_0)^a=\left(\pderivative{u}\right)^a+\frac{1}{2}(c-1)\left(\pderivative{\Omega}\right)^a,\\
		(\bm{e}_1)^a=\left(\pderivative{u}\right)^a+\frac{1}{2}(c+1)\left(\pderivative{\Omega}\right)^a,\\
		(\bm{e}_2)^a=\frac{1}{b}\left(\pderivative{\theta}\right)^a,\\
		\begin{align}
			(\bm{e}_3)^a=&\frac{a\sin\theta}{b}\left(\pderivative{u}\right)^a+\frac{a\Omega^2\sin\theta}{b}\left(\pderivative{\Omega}\right)^a\notag\\
			&+\frac{1}{b\sin\theta}\left(\pderivative{\varphi_+}\right)^a,
		\end{align}
	\end{gather}
	where
	\begin{gather}
		b=\sqrt{1+a^2\Omega^2\cos^2\theta},\\
		c=\Omega^2-\frac{2M\Omega^3}{b^2},
	\end{gather}
	it can be seen that the causal structure is determined and $\Omega=0$ is $\mathcal{I}^-$ of $C^\prime$.
	We obtain the conclusion that $\mathcal{I}^+$ of $A$ is $\mathcal{I}^-$ of $C^\prime$ as shown in Fig.~\ref{fig:3}.
	
	In inner Eddington coordinate with conformal factor $\Omega$, the conformal metric which is regular at $\mathcal{H}_+^+$ and $\mathcal{I}^-$ takes the form \cite{chandrasekhar1985mathematical},
	\begin{align}
		\mathrm{d}\tilde{s}^2=&-\left(\Omega^2-\frac{2M\Omega}{\rho^2}\right)\mathrm{d}v^2\notag\\
		&-\frac{4aM\Omega\sin^2\theta}{\rho^2}\mathrm{d}v\mathrm{d}\varphi_-\notag\\
		&+\left(1+a^2 \Omega^2+\frac{2Ma^2 \Omega\sin^2\theta}{\rho^2}\right)\sin^2\theta\mathrm{d}\varphi_-^2\notag\\
		&+2a\sin^2\theta\mathrm{d}\varphi_-\mathrm{d}\Omega
		-2\mathrm{d}v\mathrm{d}\Omega\notag\\
		&+\Omega^2\rho^2\mathrm{d}\theta^2.
	\end{align}
	\begin{figure}[t]
		\includegraphics[width=8.6cm]{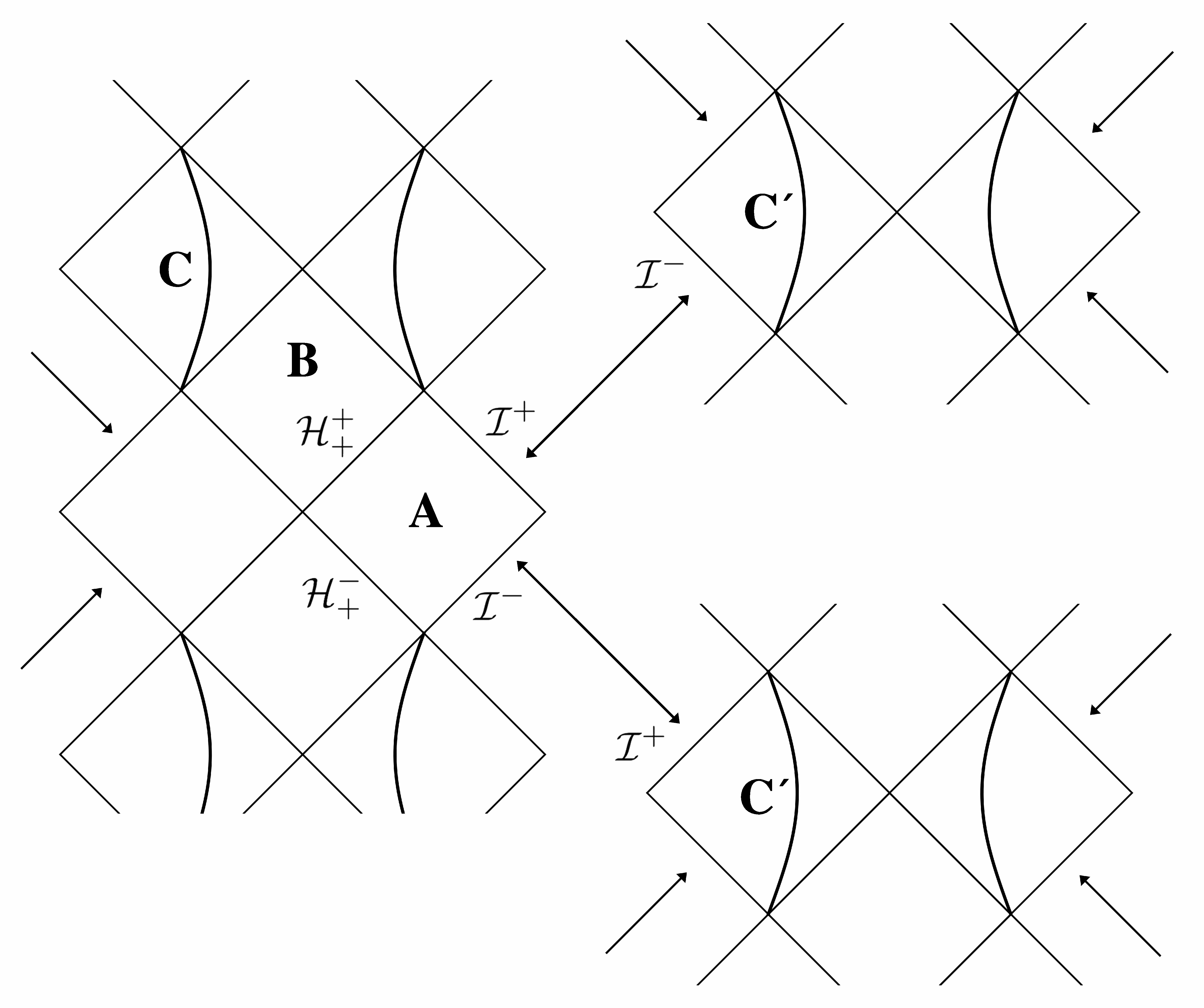}
		\caption{\label{fig:4} Conformal completion of Kerr spacetime. The bold lines represent the hypersurfaces of $r=0$. When $\theta=\pi/2$, the regions of $\bm{K}$ where $r$ is less than zero and the regions of $\bm{K^\prime}$ where $r$ is greater than zero do not exist. The arrows show that there are infinitely many related $\bm{K}s$ and $\bm{K^\prime}s$.}
	\end{figure}
	
	Similarly, we deduce that $\mathcal{I}^-$ of $A$ is just $\mathcal{I}^+$ of $C^\prime$, and $\mathcal{I}^-$ is also a killing horizon with vanishing surface gravity.
	There are three separate classes of killing horizons in the conformal extended Kerr spacetime, which makes it novel.
	
	Beyond $\mathcal{I}^+$ of $A$, there is clearly a rotating black hole $\bm{K^\prime}$.
	It is rational to extend $C^\prime$ to $r\geq0$ when $\theta\neq\pi/2$.
	Additionally, it cannot be extended and only has regions of $-\infty<\Omega<0$ when $\theta=\pi/2$ because of the existence of a singularity. 
	This makes $\bm{K^\prime}$ distinct from Kerr black hole $\bm{K}$.
	And the region beyond $\mathcal{I}^+$ of $C^\prime$ is another $\bm{K}$. 
	Also, the region beyond $\mathcal{I}^-$ of $A$ is another $\bm{K^\prime}$. 
	Therefore, the conformal extended spacetime contains infinitely many $\bm{K^\prime}s$ and $\bm{K}s$ as shown in Fig.~\ref{fig:4}.
	If the time orientation of the original $\bm{K}$ is known, then every $\bm{K}$ or $\bm{K^\prime}$'s time orientation is known.
	
	Similarly, the complete structure of the conformal extended Kerr spacetime is not unique but there is a fixed globally hyperbolic subspacetime $\{
	\bm{M}=A\cup B\cup A^\prime \cup B^\prime \cup \mathcal{H}_+,\tilde{g}_{ab}\}$ with spacelike Cauchy surfaces.
	According to the results in \cite{ottewill2000renormalized} and the same methods above, a massless scalar field $\tilde{\phi}$ on the subspacetime can be quantized, and then CCH and FT vacuums for it can be obtained.
	It can be observed that the probability that a static or staionary two-level system at ground state will be excited near $\mathcal{I}^-$ is not zero.
	Therefore $C^\prime$ is emitting thermal radiation that is traveling through $\mathcal{I}^-$.
	Given that it resembles the heat radiation coming from past $B$, this seems reasonable. It is plausible to think that conformal extended Kerr spacetime is more physical since it makes the Unruh effect make perfect sense.
	Additionally, by multiplying $\tilde{\phi}$ by $1/r$, we can restrict the quantized field to $\{\bm{M},g_{ab}\}$.
	The vacuums for the confined field are just the usual CCH and FT vacuums.
	It also implies a connection between the thermal radiation passing through$\mathcal{I}^-$ in the conformal extended spacetime and the thermal radiation coming from $\mathcal{I}^-$ in Kerr spacetime.
	
	In conclusion, we discover that Schwarzschild spacetime can be conformal extended, and a black hole $C$ with a timelike singularity is generated when an adequate conformal factor is given.
	The complete structure of the conformal extended spacetime is obtained here.
	When a quantized massless scalar field is examined on the conformal extended spacetime, there is thermal radiation coming from $C$.
	This makes sense because it is similar to the thermal radiation coming from the white hole.
	We come to the conclusion that it perfectly explains the Unruh effect.
	If we confine the field to Schwarzschild spacetime, we get the previously unimaginable result of thermal radiation originating from past null infinity.
	Furthermore, we discover that in Kerr spacetime, the conformal extension beyond a null infinity is a rotating black hole, which is different from a Kerr black hole.
	Additionally, there are an endless number of $\bm{K}s$ and $\bm{K^\prime} s$ in the conformal extended spacetime.
	When a quantized massless scalar field is examined on the conformal extended spacetime, there is thermal radiation travels through null infinity.
	In a similar vein, we conclude that it fully explains the Unruh effect.
	If we limit the field to Kerr spacetime, thermal radiation comes from past null infinity, which was unimaginable before.
	We need to interpret these issues because the combination of general relativity and quantum mechanics leads to non-physical consequences in some otherwise physical spacetimes.
	In this paper, we rationalize the Unruh effect by extending black hole spacetime and provide an interpretation of the problem of thermal radiation coming from infinity in non-extended spacetime.
	Clearly, this result is consistent in two representatively geometrically different black holes.
	Additionally, the strategy in this study could be applied to other spacetimes, like Minkowski spacetime and the Reissner-Nordstrom black hole.
	
	H. R. L. is supported by the National Science Foundation of China (Grant No. 11774284).
	
%

\end{document}